# A Review on Application of Data Mining Techniques to Combat Natural Disasters

Saptarsi Goswami[(1)]
Assistant Professor[(1)]
Institute of Engineering and Management, Kolkata, India[(1)]
saptarsi.goswami@iemcal.com[(1)]

Sanjay Chakraborty[(2)]
Assistant Professor[(2)]
Institute of Engineering and Management, Kolkata, India[(2)]
sanjay.chakraborty@iemcal.com[(2)]
Mob: 09038205310

Sanhita Ghosh[(3)]
Assistant Professor[(3)]
Institute of Engineering and Management, Kolkata, India[(3)]
sanhita.ghosh@iemcal.com[(3)]

Amlan Chakrabarti[(4)]
Associate Professor & HoD[(4)]
A.K.Choudhury School of Information Technology, Kolkata, India[(4)]
acakcs@caluniv.ac.in[(4)]

Basabi Chakraborty[(5)]
Professor[(5)]
Faculty of Software and Information Science[(5)]
Iwate Prefectural University, Japan[(5)]
basabi@iwate-pu.ac.jp[(5)]

*Abstract*

*Thousands of human lives are lost every year around the globe, apart from significant damage on property, animal life etc.due to natural disasters (e.g., earthquake, flood, tsunami, hurricane and other storms, landslides, cloudburst, heat wave, forest fire). In this paper, we focus on reviewing the application of data mining and analytical techniques designed so far for i) prediction ii) detection and iii) development of appropriate disaster management strategy based on the collected data from disasters. A detailed description of availability of data from geological observatories (seismological, hydrological), satellites, remote sensing and newer sources like social networking sites as twitter is presented. An extensive and in depth literature study on current techniques for disaster prediction, detection and management has been done and the results are summarized according to various types of disasters. Finally a framework for building a disaster management database for India hosted on open source Big Data platform like Hadoop in a phased manner has been proposed.*

## 1. Introduction

Natural disasters affect human and animal lives and properties all round the globe. In many cases the reasons are not in our control. As noted in [60], for the three decades namely1970-80 (rank 2[nd]), 1980-90 (rank 4[th]), 1990- 00 (rank 2[nd]), India ranks in first 5 countries in terms of absolute no. of the loss of human life. It's not only the immediate effect as observed in [61], exposure to a natural disaster in the past months increases the likelihood of acute illnesses such as diarrhea, fever, and acute respiratory illness in children under 5 year by 9–18%.. The socioeconomic status of the households has a direct bearing on the magnitude and nature of these effects. The disasters have pronounced effects on business houses as well. As stated in [50] 40% of the companies, which were closed for consecutive 3 days, failed or closed down within a period of 36 months. The disasters are not infrequent as well.Only for earthquake [7], there are as many as 20 earthquakes every year which has a Richter scale reading greater than 7.0. The effects of the disasters are much more pronounced in developing countries like India.

Meteorologist,Geologists, Environmental Scientists, Computer Scientistsand scientists from various other disciplines have put a lot of concerted efforts to predict the time, place and severity of the disasters. Apart from advanced weather forecasting models, data mining models also have been used for the same purpose. Another line of research, has concentrated on disaster management, appropriate flow of information, channelizing the relief work and analysis of needs or concerns of the victims. The sources of the underlying data for such tasks have often been social media and other internet media.Diverse data are also collected on regular basis by satellites, wireless and remote sensors, national meteorological and geological departments, NGOs, various other international, government and private bodies, before, during and after the disaster. The data thus collected qualifies to be called 'Big Data' because of the volume, variety and the velocity in which the data are generated.

A brief technical description of some of the major natural disasters:-

- Earthquake- A sudden movement of the earth's crust, causing destruction due to violent activities caused due to volcanic action underneath the surface of the earth. 55% of India's landmass are in seismic zone III-V.

- Landslide – A sudden collapse of the earth or mass of rock from mountains or cliff due to vibration on the earth's surface. In India the northern sub-Himalayan region and Western Ghats are prone to landslides.

- Cloudburst- It is an extreme form of unpredicted rainfall in the form of thunder storm, hail storm and heavy precipitation which is short lived. Unseasonal heavy rainfalls are common in India. A devastating effect of it was the flash flood in North India in 2013 that killed thousands of pilgrims and animals.

- Storm- A bad weather in the form of rain or snow caused by strong winds or air currents formed due to unexpected changes in air pressure on the earth's surface. Cyclones are common in various parts of India, especially the coastal regions that leave long lasting and expensive damages to human lives and properties.

- Flood- An overflow of huge water masses beyond normal limits over dry land. Every year, millions of human lives, cattle and agricultural crops are destroyed in India due to lack of planning and improper weather forecasting.

- Tsunami- High sea waves that are large volumes of displaced water, caused due to an earthquake, volcanic eruption or any other underwater explosions. The 2004 Tsunami that hit parts of the southeastern coast of India had devastating effects on the mainland and Andaman and Nicobar Islands.

- Volcanic Eruption- It is a sudden, violent discharge of steam, gases, ashes, molten rocks or lava from the surface of the earth that are ejected to heights and spread for several miles. Underwater volcanoes and on the islands surrounding the landmass of India are common. However, they have not imposed significant damages to the mainland till now.

The unique contributions of the paper are as follows: -

- ✓ A comprehensive summary of different data mining techniques applied to various taskspertaining to the natural disasters
- ✓ A detailed account of various types and sources of data for each category of task and

disaster
- ✓ A brief account of disaster management 'status-quo' from Indian context
- ✓ A brief review of suitability of 'Twitter' as a data source
- ✓ A presentation of proposed architecture to streamline disaster management

The organization of the paper is as follows: In Section II, natural disasters have been discussed with focus on India; a brief description of the existing disaster management structure is also outlined. In Section III, the broad categorization of the tasks that can be achieved with respect to natural disaster are presented in detail. In Section IV, granularlevels of tasks are enlisted with respect to the major type of tasks discussed in Section III. Details of the tasks, data used data mining methods used; country orregion hasalso been discussed. In Section V, a structured view of different types of required data and their corresponding sources has been discussed. A short review of Twitter and other Internet resources as data source has been discussed, along with their application for natural disaster in Section VI. In Section VII, a process flow and architecture of a disaster management system has been proposed. Section VIII contains, conclusionwith the direction of future work.

## 2. NATURAL DISASTER FROM AN INDIAN CONTEXT

India is vulnerable to various natural disasters due to its unique geo-climatic condition as a result of its geographical location. This subcontinent is surrounded by water bodies on three sides and the Himalayas on the North. The country has been hit approximately by 8 natural calamities per year and there has been about 5 times increase in frequency of natural disasters in the past three decades. The calamities that affect the country can be categorized as :- 57% landmass is prone to earthquakes, 12% floods (about 40 million hectares of land is vulnerable to floods) and 8% are prone to cyclones. Table 1 , records such disasters for last 15 years.

**Table 1: Natural disasters in India in last 15 years**

| Disaster Type | Year | Origin (India) | Tolls |
|---|---|---|---|
| Earthquake | 2001 | Gujrat | 20,000 |
|  | 1999 | Chamoli | 150 |
| Cyclones | 2012 | Tamil Nadu | 20 |
|  | 2011 | Tamil Nadu | 41 |
|  | 2010 | Andhra Pradesh | 32 |
|  | 2009 | West Bengal | 100 |
|  | 1999 | Orissa | 15,000 |
| Tsunami | 2004 | Indian Ocean | 230,000 |
| Floods | 2007 | Bihar | 41 |
|  | 2005 | Mumbai | 5,000 |
| Cloud Burst | 2014 | Jammu & Kashmir | 4,500 |
|  | 2013 | Uttarakhand | 5,700 |
| Landslides | 2014 | Manlin, Pune | 28 |
|  | 1998 | Malpa, Manasarovar Yatra | 380 |

Asia, tops in terms of no. of disaster events among the continents. Close to 60% of the disasters in Asia are originated in south Asia and 40% are originated in India.In the below figure ( Figure 1), the above statistics are displayed.

Figure 1: Disaster Trends across the globe.

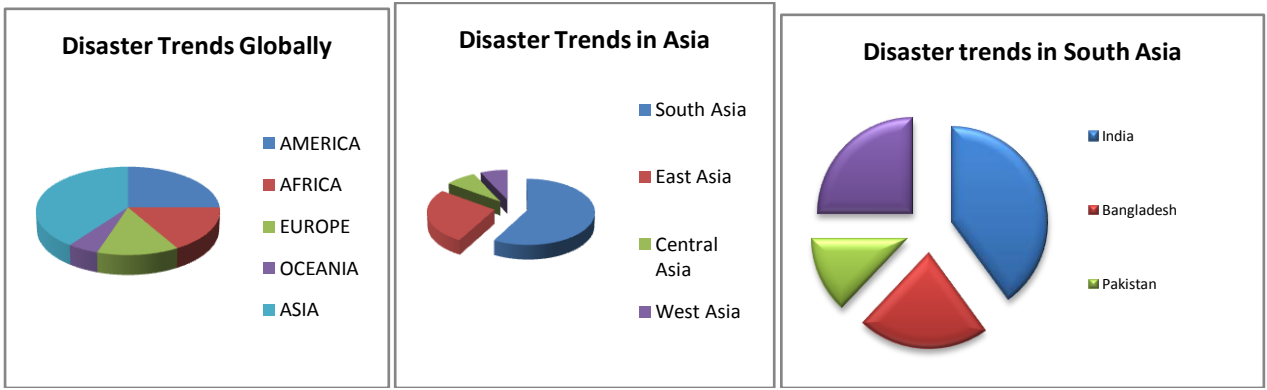

India is a victim of natural disasters every year and the loss of lives and properties add up to millions of rupees which this country cannot afford to lose. There are certain reasons for such poor disaster management procedures followed in this country.
- Inadequate early warning system
- Poor preparation before the disaster occurs
- Inadequate and slow relief operation
- Lack of proper administration
- Slow process of rehabilitation and reconstruction
- Poor management of finances for relief work
- Lack of effective help to victims

The apex body that handles disaster management in India is the National Disaster Management Authority (NDMA) whose Chairman is the Prime Minister himself. Similar authorities are also set up at state and district levels which are respectively headed by the Chief Ministers and Collectors or Zilla Parishad Chairperson. The Natural Urban Renewal Mission has been set up in 70 cities due to the recent unprecedented weather conditions in major metros and megacities. Need of research to predict, prevent and reduce means of losses from a disaster is far from over. Over 100,000 Rural Knowledge Centers (or IT Kiosks) have to be established for meeting the need for spatial E-Governance and therefore offering informed decisions in disaster prone areas to improve the response time and type of relief aid offered to the victims on time. National disaster management structure is depicted in Figure 2.

Figure 2: Disaster management structure India.

## National Disaster Management Structure

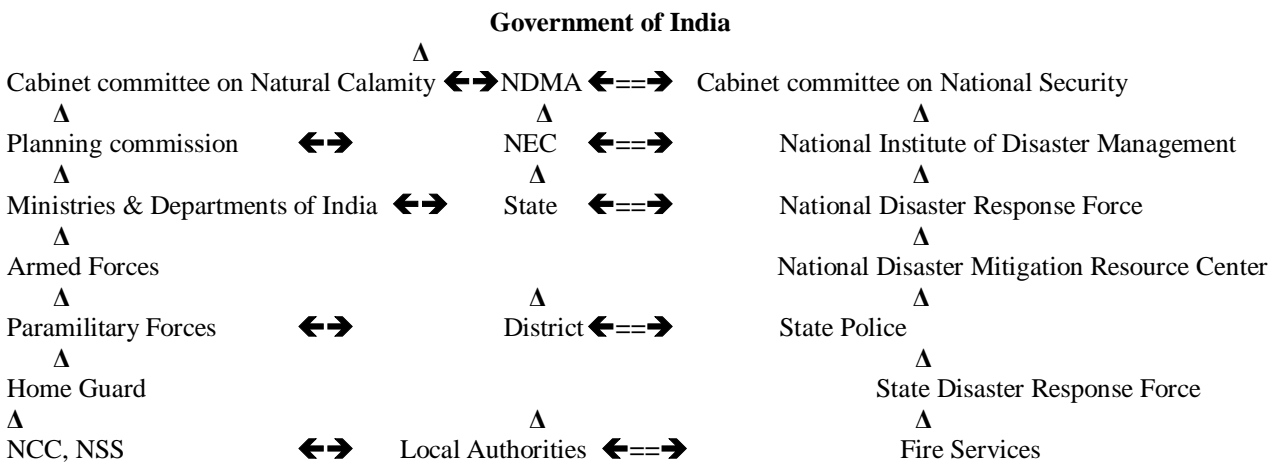

There are some model agencies in India who are responsible for Disaster Management

- Floods → Ministry of Water Resource CWC
- Cyclones → Indian Meteorological Department

Ministry of Agriculture and Ministry of Animal Husbandry are involved in all the above

## 3. BROAD CATEGORY OF TASKS WITH RESPECT TO NATURAL DISASTERS

In this section, broad categories of tasks that can be solved using different types of datahave been discussed. We can classify the objectives of the tasks, in the following three major categories

1. Prediction: - These sets of tasks involve prediction of the natural disaster, disaster prone area and different attributes of a natural disaster that can occur. Basically these tasks involve prediction or forecasting of time, place and magnitude of the disaster.

2. Detection: - These sets of tasks involve detection of the natural disaster promptly after it has occurred. Literature studies indicate that the social sensors in terms of tweets, other social media websites report a natural disaster much faster than the observatories.

3. Disaster Management Strategies: These methods deal with identification of different entities that are taking part in combating a disaster so that communication is enhanced, appropriate concern of the affected people are identified and distribution of relief items are optimized.

Another branch of study deals with carrying out the psychological and behavioral changes over affected regions after the disaster.

In many cases, the classification stated above is overlapping. As an example, it can be surely argued that, detecting the natural disaster helps in disaster management strategies. Even the psychological studies can give lot of insight to disaster management strategies. So the classification is based on the direct objective of the task involved. There are some rare cases which fall in a borderline.

**Prediction:** There is no doubt that this would be the most 'ideal' problem to solve. But very often, this is not a problem that can be solved with available data and techniques. However, it is possible to predict, the areas which are susceptible to a particular type of disaster, let's say, landslide or flood. The prediction techniques have been seen to be of more use for predicting various characteristics of a natural disaster, which has occurred. As an example, the techniques can be used to predict the magnitude of an earthquake, track and intensity of a cyclone etc. Analysis of various spatial and temporal data is often needed for such tasks. Though handful, another branch of research has focused on using unusual animal behavior to predict a natural disaster.

**Detection:** Often the meteorological observatories detect the natural disaster, but the news of the detection takes a long time to be communicated to proper authorities with the exact location of the detection.

**Disaster Management Strategies:** These sets of tasks are involved in forming appropriate disaster management strategies. An example of such tasks is identifying critical entities for disaster management, identify proper communication study, identifying the needs of the disaster affected area. Social media data is very important in these types of tasks.

The aim of disaster management should be the following-
- Minimize casualties
- Rescue victims on time
- Offer first aid instantly
- Evacuate people and animals to safe places
- Reconstruct the damages immediately

## 4. DATA, MODELS, TASKS

In this section, a summary has been enclosed highlighting the granular level tasks corresponding to the major type of tasks like prediction, detection and disaster management. An account of model /techniques has been given along with the data used and the country of the disaster / research.
The findings are summarized from table 2a to 2h.The tables are divided as per the natural disasters; separate tables are presented for earthquake, cloudburst, landslide, flood, storm, tsunami and volcanic eruption. In the last table, that is table 2h,generic efforts without focus on a specific type of disaster have been covered.

**Table 2a: Earthquake data, model and task summary**

| Task | Detailed Objective | Model techniques Used | Data Source, type | Country |
|---|---|---|---|---|
| **Prediction** | Predict magnitude of earthquake[2] | Particle Swarm Optimization | Seismological Data | China |
| | Focus on abnormal animal behavior, rather than the geophysical indicators. The study has been done mainly in Japan, China and USA. Animals are much more sensitive to the change in electric field precursor to the earthquake. [5] | NA | NA | Japan, China, USA |
| | Predict magnitude of earthquake [6] | Neural Network | Seismological Data | USA |
| | Building a data warehouse for earthquakes, for uniformity of data and structure of data for a uniform interchange and better decision making. [7] | Ontology, Star Schema, Data Warehouse | Seismological Data | All over the world |
| | Predict earthquake based on time series data. [8] | Non Linear Time Series and Fuzzy Rules | Seismological Data | All over the world |
| | Predict earthquake from historical data and also propose a grid system for distributed processing and better information interchange. [9] | Feature Generation and Clustering | Seismological & GIS Data | USA |
| **Detection** | Discover major earthquakes faster than seismological observatories [1] | Text Mining | Twitter | USA |
| | The affected area citizens visit web pages of the Swiss Seismological Service, By doing an IP tracing and volume analysis, the affected regions can be tracked easily [3] | Regular log mining techniques | Web Server logs | Switzerland |
| | To detect earthquake from Social Sensors i.e. Twitter, do a Spatial, temporal analysis and send notification much faster than that of the Japan Meteorological Agency (JMA) [11] | Temporal Analysis, Kalman Filter | Twitter | Japan |
| **Disaster Management** | A temporal analysis of peoples need after the earthquake from blogs and social media. This can make the relief operation more effective [4] | Text Mining, latent Semantic Analysis (LSA), Time Series | Blogs & Social Media Data | Japan |
| **Behavioral and social analysis** | To study general peoples reaction after a natural disaster like an earthquake and how long they take to subside to normal level. [10] | Time Series, Text Processing | Twitter | Japan |

Table 2b: Cloudburst data, model and task summary

| Task | Detailed Objective | Model techniques Used | Data Source, type | Country |
|---|---|---|---|---|
| **Prediction** | Observe different parameters of climate from the earth science data, to find out if there was enough indication of the Uttarakhand disaster. [12] | Anomaly Detection, Time Series | Earth Science Data | India |
| | To leverage OLAP Structure to store metrological data and analyze them to identify cloudbursts [13] | OLAP Cubes, K Means clustering | Meteorological data | India |
| | Real-time newscast and prediction of rainfall in case of extreme weather like cloud burst from Doppler weather radar data [14] | Mesoscale Model | Doppler Weather Radar data (DWR) | India |

Table 2c: Flood data, model and task summary

| Task | Detailed Objective | Model techniques Used | Data Source, type | Country |
|---|---|---|---|---|
| **Prediction** | Build a model and select appropriate | Decision Tree | Hydrological data, | Germany |

| | parameters to assess the damage from flood [16] | | Remote sensing data , GIS | |
|---|---|---|---|---|
| | To build a model to find susceptible flood regions based on spatial data [17] | Logistic Regression and Frequency Ratio Model | Meteorological data ( digital elevation model) , river , rainfall data etc. | Malaysia |
| | To build a model to predict monsoon flood (1 day ahead). The built system gave better results than existing auto regressive models. [19] | Wavelet Transform, Genetic Algorithm, Artificial neural net | Hydrological time series data | India |
| | Build a system for flood forecast for medium- to large-scale African river basins ( Before 2 Weeks) [18] | Probabilistic Model & Ensemble | Hydrological data | Africa |
| | Build a Flood routing model based on past data.[20] | Muskingum flood routing model, Cuckoo Search (For parameter values and calibration) | Hydrological and Hydraulic Data | All World |
| **Disaster Management** | A study of tweets during various floods was done to identify key players. The study shows the effect of local authority involvement in successfully tackling a disaster.[15] | Text Mining Methods | Twitter | Australia |

**Table 2d: Landslide data, model and task summary**

| Task | Detailed Objective | Model techniques Used | Data Source, type | Country |
|---|---|---|---|---|
| **Prediction** | Build a classifier to identify landscapes to landslide susceptible areas based on soil properties, geomorphological, and groundwater conditions etc. [21] | Discrete Rough Set and C4.5 Decision Tree | Remote Sensing Data and GIS | Taiwan |
| | To build a classification model to predict land slide. The various factors considered are rainfall, land use, soil type, slope etc.[22] | SVM, Naïve Bayes | GIS (rainfall, land use, soil type, slope and its) | India |
| | A generic note on usefulness of data mining/machine learning models in predicting place and time of a land slide[23] | NA | NA | All over the world |
| | Build a prediction model based on an inexpensive wireless sensors placed on susceptible regions. [24] | Distributes statistical prediction method | Wireless Sensor Data | India |
| | To build a model to identify areas of shallow landslide. [25] | Spatial Distribution | Geomorphologic information and hydrological records | Taiwan |
| | Predicting landslide based on past data [41] | Back propagation Neural Network, Genetic Algorithm, Simulated Annealing | | China |

**Table 2e: Volcanic Eruption data, model and task summary**

| Task | Detailed Objective | Model techniques Used | Data Source, type | Country |
|---|---|---|---|---|
| **Prediction** | Analysis of multivariate time series data to understand the state of the volcano and | Multivariate Time Series clustering | Geophysical data through | Italy |

| | potential hazard assessment [26] | | monitoring network | |
|---|---|---|---|---|
| | To monitor and predict trajectories of volcanic ash cloud, to minimize air crash [39] | Not mentioned | Plume height, mass eruption rate, eruption duration, ash distribution with altitude, and grain-size distribution | USA |

**Table 2f: Storm data, model and task summary**

| Task | Detailed Objective | Model techniques Used | Data Source, type | Country |
|---|---|---|---|---|
| **Prediction** | Take the data of current storm and compare with the historical & synthetic storms using storm similarity index (SSI) from the databases to understand the effect. Visualization of the storm path is done on Google Earth. The study was done on two previous storms Katrina and Camille. [28] | Data Mining Techniques | National Hurricane Center (NHC) | USA |
| | To predict Cyclone Track data for coming 24 hours, based on past 12-hour locations at six hourly intervals besides the present position about the latitude and longitude[29] | Artificial Neural Network (ANN) | 32 Years Tropical Cyclone data on Indian Ocean from Joint Typhoon Warning Center (JTWC), USA | India |
| | Detect storm surge using no linear model from data collected at coastal station[32] | Time series and chaos theory | Water level, surge, atmospheric pressure and wind speed/direction data from seven coastal stations along the Dutch coast are monitored and provided by the North Sea Directorate | Netherland |
| **Behavioral** | Study effect of the hurricane 'Hugo' on life events birth, death, divorce etc. [30] | Statistical analysis | Life event data from all the counties of South Carolina | USA |
| **Disaster management** | Identify the concerns of people, stay duration of the 'concerns', conductanalysis by gender [31] | Sentiment analysis, normal text processing Techniques | Twitter | USA |
| | Analysis of people's sentiment, after Hurricane Sandy and also to gather and decimate important information through social media. [27] | Text processing techniques | Twitter | USA |
| | Analysis of public behaviors during and after a disaster through visualization and spatial temporal analysis [59] | Spatial, temporal techniques, visualization | Twitter | USA |

**Table 2g: Tsunami data, model and task summary**

| Task | Detailed Objective | Model | Data Source, | Country |
|---|---|---|---|---|

|  |  | techniques Used | type |  |
|---|---|---|---|---|
| **Disaster Management** | Viability of use of twitters by government agencies, to inform public about natural disaster. It was compared against traditional sources and proved its value as a complementary source. [33], [35] | Text processing techniques | Twitter | Indonesia, USA |
| **Prediction** | Build an early warning system for Tsunami. [36] | Flood filling algorithm, classification algorithms | Bathymetry data, Seismic data, Sea wave conditions, web service and API to collect the data | Indonesia |

**Table 2h: General data, model and task summary**

| Task | Detailed Objective | Model techniques Used | Data Source, type | Country |
|---|---|---|---|---|
| **Disaster Management** | Build a tool to extract important information from tweets for relief workers. [37] | Text processing techniques | Twitter | USA |
|  | Build a system, for disaster discovery and humanitarian relief based on tweets. The system consists of a stream reader, a data storage and a visualization module. [38] | SVM, LDA, Topic clustering | Twitter | USA |
| **Prediction** | Build a geo hazard database for early prediction system, by using the Google news service. Geo tagging is done for geo referencing. The purpose is using this database extensively for disaster management. [40] | Text processing | Google News Service, RSS Feed | Italy |

In the above tables, different research directions in combating natural disaster have been discussed. This is a multidisciplinary activity needing experts from environmental science, geology, meteorology, social science, computer science etc. The above list is not exhaustive, but an effort has been made to cover last 10 -12 years data in this section. Here are few of our observations

- Twitter as a source has become important for real time detection and understanding of the need and concern of the affected people.10 out of the 40 papers we reviewed above use twitter as the data source. Interestingly, we did not find any referential work, where twitter has been used in an Indian context.
- It's also observed, though India has a much higher loss in terms of human life and property,adequate research as in countries like USA has not been done in India (Figure 3).

Figure 3: # Papersby country of analysis

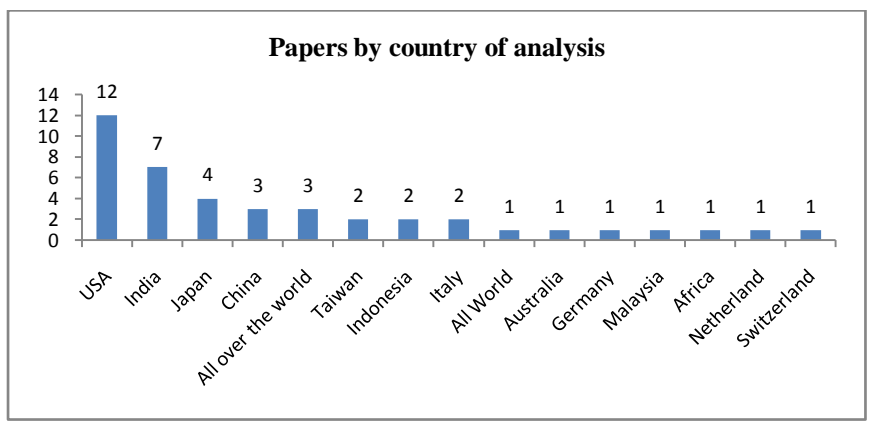

- Level of activity, in the research communities around each disaster type can be roughly estimated by Figure 4. We have used the no. of results from GoogleScholar, using the disaster specific keyword. We have restricted the search results using the 'since 2014' filter. The above searches were not limited to the application of data mining.

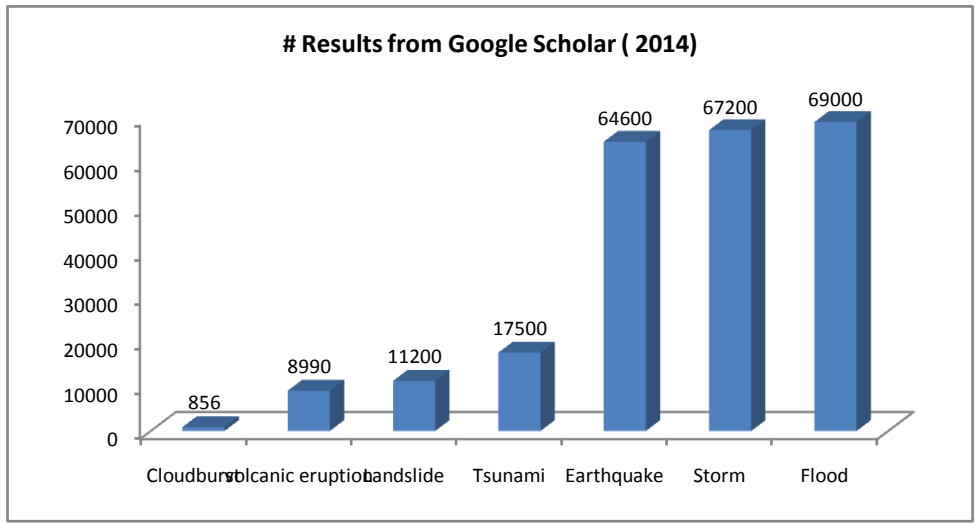

Figure 4: # Results from Google Scholar

- The main tasks, where research activitieshave been going on are prediction and disaster management respectively. As expected, the prediction tasks for each disaster are varied. Apart from the data mining tasks, quite a few papers [7,13,40] focused on building a data warehouse and OLAP structures for better information decimation and consumption. Below in Table 3,a summary of the prediction tasks that are being researched for each disaster type has been presented.

Table 3: Prediction tasks for each disaster type

| Disaster Type | Prediction Task |
|---|---|
| Earthquake | Predicting the time , place , magnitude of the earthquake |
| Cloudburst | Predicting cloudburst, predicting amount of rainfall |
| Strom | Predict the track and wind speed of the storm |
| Flood | Identify flood susceptible areas, predict flood, build flood routing model , build a model to assess damage to property etc. due to flood. |
| Landslide | Predicting landslide, predicting landslide susceptible areas. |
| Volcanic Eruption | Predict eruptions; predict the trajectory of the ash cloud |
| Tsunami | Build early warning system with tsunami |

- We find Neural Network, SVM, Decision Tree etc. have been used extensively as the data mining models. As many of the data are actually time series, time domain techniques as well as frequency domain techniques like wavelet transformation have been used. Evolutionary techniques like Genetic Algorithm and Particle Swarm Algorithm have also been used. For newer sources like Twitters, Blogs, Server Logs, News text processing techniques (Topic clustering, LDA) have been applied. Word cloud visualization has been applied on the techniques and it is shown in Figure 5. Though most of the data obtained are high dimensional in nature, apart from LDA or wavelet transform, we do not see uses of feature selection or dimensionality reduction in the approaches.

Figure 5: Word Cloud of data mining techniques on disaster management

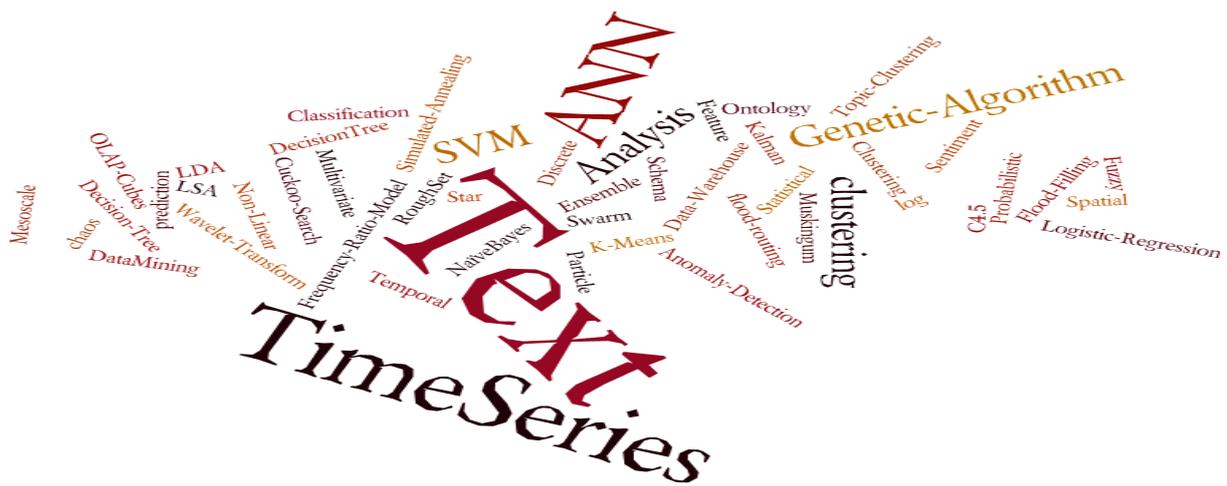

## 5. DATA SOURCES &TYPES

The objective of this section is to give researchers and practitioners a high level overview of the type of data that are useful for analysis and prediction of a natural disaster. Much of the data will qualify to be called 'Big Data', because of all or some of the dimensions of volume, variety and velocity as listed below:-
1. Volume (GIS Data, Meteorological Data, Social Media Data)
2. Variety (Text, Time Series, Spatial Data, GIS Images)
3. Velocity (because of the rate in which data is generated as well as because of the speed in which a decision needs to be taken)

We try to answer the following questions in this section:-
- What are the different types of data that are useful for each type of disaster?
- What are the sources and format of such data, at national and international level?
- How the data can be accessed? Whether it is freely available or not?

In Table 4, we have listed data types for different disaster types, their corresponding data types, the format of the data and the various agencies that collect or capture the data. The abbreviations of the agencies are used in Table 4, the details of the agencies with a URL is available in the appendix

**Table 4: Data Types and Sources of data related to disasters.**

| Disaster Types | Data Category | Data Format | Types of Data | Agencies/Bodies | Availability (Paid/Free) |
|---|---|---|---|---|---|
| Earthquakes, Tsunami etc. | Seismological Data | Magnitude of richter scale & Microsoft Excel sheet | small scale (regional) seismic macro zona-tion at scales 1:5,000,000 to 1:50,000, and large scale (local) seismic micro zonation at scales of 1:50-25,000 to 1:10,000. | ISC, NEIC, IMD, IRIS, NGDC, NWS, USGS, EM-DAT, JMA, NDMA. | Paid |
| Tsunami, Flood, Drought etc. | Hydrological Data | Google Earth Image data | Water level, Pressure & Density of water etc. | INCOIS, NOAA, EM-DAT, JMA, NDMA. | Paid |
| Landslide | Geological Data | Different devices have different storage formats. | Rainfall, Moisture, Pore pressure, Tilt, Vibrations etc. | USGS, NOAA, NODC. | Paid |
| All types of Disasters. | Remote Sensing Data | Mainly image file formats. | Spatial, temporal, and thematic data (Satellite Data). | CRSSP, NASA, ISRO, NRSC, | Open source |
| All types of Disasters. | Google Information System (GIS) & Google Mapping Data | Stored into Microsoft Excel spreadsheets or text file. | Spatial, temporal, and thematic data (Floods and pre-flood SAR images) are Collected. | NOAA, NIDM, NASA, ISRO. | Non-Paid (Sometimes Copyright assertions) |
| Flood, | Wireless Sensor | Transmitting and | Form of Analog | GSI. | Mainly from |

| Tsunami etc. | Network (WSN) Data | receiving data through wireless transmitter and receiver. | warning signal. [Needed an Analog to Digital Converter] | | Geological Survey of India (GSI) and Paid. |
|---|---|---|---|---|---|
| Volcanic Eruption | Geological Data&Seismological Data | Magnitude of richter scale& MySQL and xml format (eg. WOVOml). | Spatial, temporal, and thematic data (eg. Angle, EDM, GPS, inSAR image, Gravity, Magnetic field etc.-WOVOdat1.1 document) | WOVOdat, JMA, DATAGOV, NGDC, USGS. | Open source – freely available. |

We would not gointo details of the data acquisition methodology, several literatures are available on the same, However we thought we should briefly mention 'flash flood' technology because of the recentness of the method.

**Flash Floods with advanced WSN Technology:** Nowadays, with the advancement of several technologies, flash flood is going to be introduced with high-tech WSN activity. According to this concept, there is a Wireless sensor device associated with a flash flood technology which is emerged into a certain level of water. If the water level reached to a certain level of threshold value, then the WS device sends a danger signal to receiver station through broadcasting. This technique is planned to be introduced in some areas of Delhi city in India [42].

Figure 6: Flash Floods with advanced WSN Technology

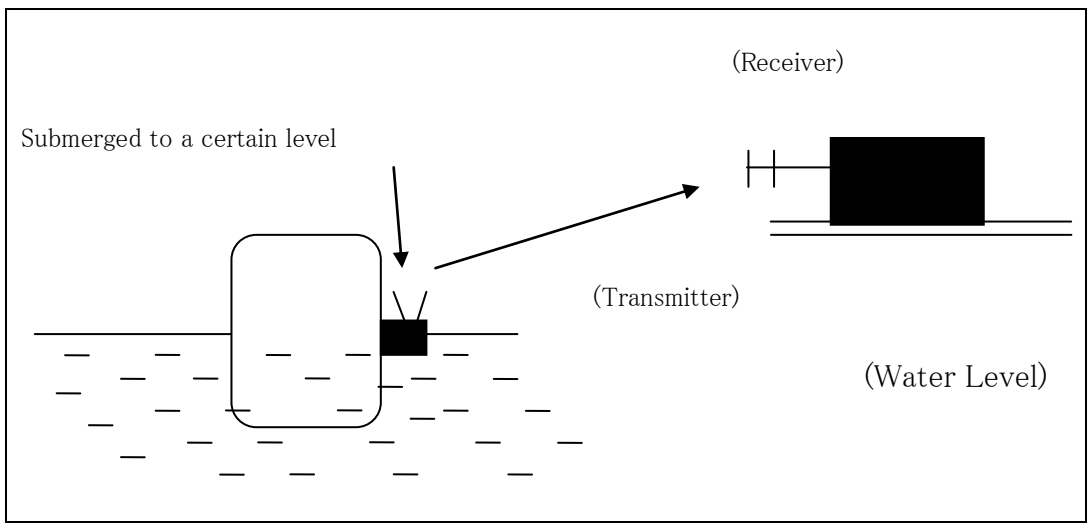

Numerical weather prediction (NWP) models, which apply advanced mathematical modeling, have been used for short term forecasting for long time. These models are employed to solve a closed set of atmospheric equations. Most of the meteorological departments have adapted using this model. However, the actual events cannot be predicted from the NWP models directly. Some other statistical techniques are required for prediction. In the paper [52], the authors list down applicability of data mining models in many weather prediction tasks.

## 6. TWITTER AND SOCIAL MEDIA AS SOURCES

Twitter as a data source has gained lot of prominence in recent years. It is ranked as one of the top 10 popular websites, having 400 million registered users and over 500 million tweets generated every day [43]. Additionally, information about disasters can be extracted from news channels & blogs through APIs, RSS feed or web scraping. Sentiment Analysis [53],[54], Stock Market [55], Public Health [56], General public mood and finding political alignments [57],[58] are some of the areas where twitter data have been used,

Some of the advantages of tweets are

- Though it is unstructured , it has some structure by the limitation of 140 characters
- It can use hashtags , which gives semantic annotations of the tweets
- The tweets have geocodes, which can help us spatially map the sources of the tweets.

Following are the fields that are available from Twitter

- **archivesource:** API source of the tweet (twitter–search or twitter–stream)
- **text:** contents of the tweet itself, in 140 characters or less
- **to_user_id:** numerical ID of the tweet recipient (for @replies) *(not always set, even for tweets containing @replies)*
- **from_user:** screen name of the tweet sender
- **id:** numerical ID of the tweet itself
- **from_user_id:** numerical ID of the tweet sender
- **iso_language_code:** code (*e.g.* en, de, fr, ...) of the sender's default language *(not necessarily matching the language of the tweet itself)*
- **source:** name or URL of the tool used for tweeting (*e.g.*, Tweetdeck, ...)
- **profile_image_url:** URL of the tweet sender's profile picture
- **geo_type:** form in which the sender's geographical coordinates are provided
- **geo_coordinates_0:** first element of the geographical coordinates
- **geo_coordinates_1:** second element of the geographical coordinates
- **created_at:** tweet timestamp in human–readable format *(set by the tweeting client — inconsistent formatting)*
- **time:** tweet timestamp as a numerical Unix timestamp

In many countries, Twitter has been used to effectively manage disasters; however in Indian context we have not seen a lot of referential work. Twitter has been effectively deployed to:-

- ✓ Detect disasters [1], [11] faster than observatories .
- ✓ Identifying key entities in disaster management& relief organization .[35], [37],[38],[15],[33]
- ✓ Temporal study of needs and concerns after the disaster [27],[31],[51]

One of the limitations of using Twitter data , is that the 'free access' (Streaming API) only provides 1% of sample data , on the other hand, the alternativeway (firehose) which provides full access is prohibitively expensive . In [43] , the authors conducted a study , between both these ways of extraction. The results obtained therein, though there is some agreement between them, to get a truer picture, the coverage of sampling in terms of parameters of steaming API , need to be varied.

## 7. PROPOSED SYSTEM

We intend to build a database for natural disasters happening in India. In, Phase 1 of the project we want to concentrate on sources like 1) Twitter 2) News 3) Other social media and Internet sources.

Like any standard systems, we will need the following components as shown in figure 7

Figure 7: Process flow of the proposed system

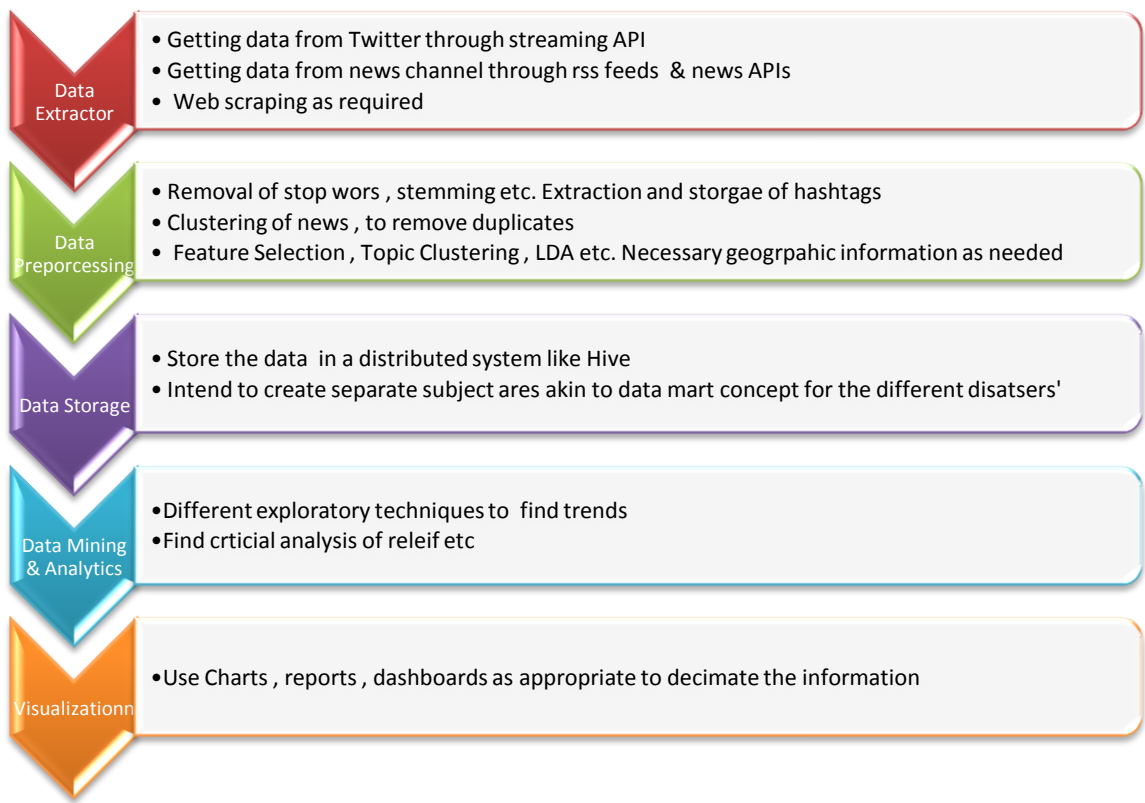

TweetTracker [38] and Tweedr [37] are couple of systems built in USA to streamline relief and disaster response. In [40], authors have proposed a geo hazards inventory store, with focus on geo tagging and entity resolution from news service. They have built a detailed database of geographical feature (Listing of Mountains, Rivers etc.) for Italy.

Our proposal is different from the above approaches, in the following ways: -

- ✓ We are targeting much wider sources, not only twitter
- ✓ The focus would be, on understanding peoples need during the disaster and evaluate the social impact and changes due to natural hazards

In extracting information we plan to implement methods as described in [44]

- In Phase 2, of our proposed system, we intend to use other sources of data apart from Internet based sources.

In Figure 8, a Hadoop based open source system has been proposed.

**Data Sources:** We are focusing on Twitter and RSS Feeds of News at this point. The data extraction scripts will be written in Flume, Scoop or R [45] as applicable.
**Data Storage:** The data storage is envisaged to be on Hadoop [46]. After data processing using 'pig' [47] the data can reside in HDFS (Hadoop distributed file system) or Hive [48], a NOSQL based database.
**Model Building:** The plan is to build our algorithms based on Mahout [49] which can leverage the HDFS and has extensive text processing capabilities and can be extended using Java.

Some portions of the proposed system is shown with dashed lines as, we plan to integrate those sources in next phases.

Figure 8: High level architecture of the proposed system

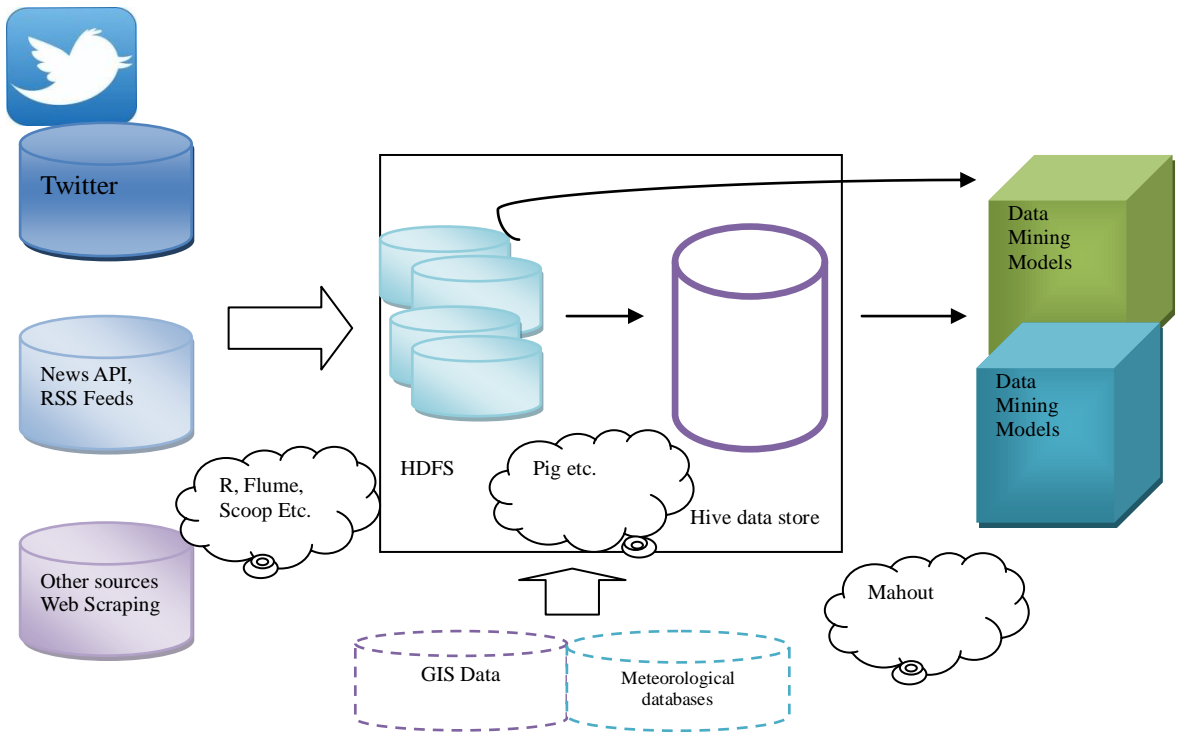

## 8. CONCLUSION

Natural disasters in forms of earthquake, floods, landslides, storms claim numerous lives,cause significant damage to property. The effects have been much more severe in a developing country like India compared to developed countries. There have been many efforts to predict the disasters based on various sources of data. In our literature survey, we explore the multidisciplinary nature of the task, where data mining models are being applied on various types of data, requiring deep subject matter expertise. Recently Social Media and Internet have also emerged as an important source of information. These sources may not be used in prediction of the disasters, but they have contributed significantly to early detection and adoption for appropriate disaster response. We observe in our study that, therehave not been enough works done in this area to tap the potential of these sourcesespecially in context of India. We propose to build a data store for natural disasters from these sources in Phase 1. In Phase II, we intend to integrate it with other sources of Information.

health and investments in rural India." Social Science & Medicine 76 (2013): 83-91.

Appendix:

| Data Agency | URL |
|---|---|
| ISC- International Seismological Center | http://www.isc.ac.uk/standards/datacollection/ |
| NEIC- National Earthquake Information Center | http://earthquake.usgs.gov/regional/neic/ |
| IMD- India Meteorological Department | http://www.imd.gov.in/ |
| IRIS-Global Seismographic Network | http://www.iris.edu/ |
| USGS- United State Geological Survey | http://www.usgs.gov/ |
| SMA- Social Media Analytic | http://www.datalabs.com.au/ |
| NODC- National Oceanographic Data Center | http://www.nodc.noaa.gov/ |
| CRSSP- Commercial Remote Sensing Space Policy | http://crssp.usgs.gov/ |
| NRSC- National Remote Sensing Center | http://www.nrsc.gov.in |
| EM-DAT (CRED) | http://www.emdat.be/additional-disaster-data-resources |
| WOVOdat – A database of Volcanic unrest | http://www.wovodat.org/ |
| JMA- Japan Meteorological Agency | http://www.jma.go.jp/jma/en/Activities/earthquake.html |
| DATAGOV | https://catalog.data.gov/dataset/global |
| NDMA- National Disaster Management Authority | http://www.ndma.gov.in/ |